\documentclass[twocolumn]{aastex63}
\usepackage{amsmath}
\usepackage{natbib}
\newcommand{\Qs}{Q_{\ast}'}
\newcommand{\logQs}{\log_{10}{Q_{\ast}'}}
\received{May 8, 2020}
\revised{July 10, 2020}
\accepted{July 16, 2020}
\submitjournal{AAS Journals}

\shorttitle{Ultra-short-period Planets are Stable Against Tidal Inspiral}
\shortauthors{Hamer \& Schlaufman}

\begin{document}

\title{Ultra-short-period Planets are Stable Against Tidal Inspiral}

\correspondingauthor{Jacob H. Hamer}
\email{jhamer3@jhu.edu}

\author[0000-0002-7993-4214]{Jacob H. Hamer}
\affiliation{Department of Physics and Astronomy, Johns Hopkins
University, 3400 N Charles St, Baltimore, MD 21218, USA}

\author[0000-0001-5761-6779]{Kevin C. Schlaufman}
\affiliation{Department of Physics and Astronomy, Johns Hopkins
University, 3400 N Charles St, Baltimore, MD 21218, USA}

\begin{abstract}

\noindent
It has been unambiguously shown in both individual systems and at the
population level that hot Jupiters experience tidal inspiral before the
end of their host stars' main sequence lifetimes.  Ultra-short-period
(USP) planets have orbital periods $P < 1$ day, rocky compositions,
and are expected to experience tidal decay on similar timescales to
hot Jupiters if the efficiency of tidal dissipation inside their host
stars parameterized as $\Qs$ is independent of $P$ and/or secondary mass
$M_{\mathrm{p}}$.  Any difference between the two classes of systems would
reveal that a constant  $\Qs$ model is insufficient.  If USP planets
experience tidal inspiral, then USP planet systems will be relatively
young compared to similar stars without USP planets.  Because it is a
proxy for relative age, we calculate the Galactic velocity dispersions of
USP planet candidate host and non-host stars using data from Gaia Data
Release 2 supplemented with ground-based radial velocities.  We find
that main sequence USP planet candidate host stars have kinematics
consistent with similar stars in the Kepler field without observed
USP planets.  This indicates that USP planet hosts have similar ages
as field stars and that USP planets do not experience tidal inspiral
during the main sequence lifetimes of their host stars.  The survival
of USP planets requires that $\Qs\gtrsim10^7$ at $P\approx0.7$ day and
$M_{\mathrm{p}}\approx2.6~M_{\Earth}$.  This result demands that $\Qs$
depend on the orbital period and/or mass of the secondary in the range
$0.5\mathrm{~days}\lesssim P\lesssim5$ days and $1~M_{\oplus}\lesssim
M_{\mathrm{p}}\lesssim1000~M_{\oplus}$.

\end{abstract}


\section{Introduction}

\citet{Hamer2019} demonstrated unambiguously at the population level that
hot Jupiters are destroyed by tides during the main sequence lifetimes of
their host stars.  Soon after, \citet{Yee2020} showed that the departure
from a linear ephemeris in the WASP-12 system could only be explained by
tidal decay.  These discoveries ended 25 years of uncertainty regarding
the stability of close-in giant planets against orbital decay due to
tidal interactions with their host stars.

With orbital periods $P < 1$ day, ultra-short-period (USP) planets are
an even more extreme population than hot Jupiters.  CoRoT-7 b, the first
transiting terrestrial exoplanet to be detected, is a USP planet with
$P=0.85$ days \citep{Leger2009}.  It was subsequently shown that CoRoT-7
is a multiple-planet system \citep{Queloz2009}.  Following this discovery,
\citet{Schlaufman2010} explained the existence of CoRoT-7-like systems
as a consequence of convergent Type I migration in multiple-planet
systems which is terminated at their parent protoplanetary disks'
magnetospheric truncation radii.  Following disk dissipation, secular
interactions between planets maintain non-zero eccentricities in a
system's innermost planet, thereby causing the orbital decay of that
planet due to tidal dissipation within it as circularization occurs.
This process continues until the innermost planet secularly decouples
from the rest of the planets in a system, halting its inward drift at
$P < 1$ day.  They argued that a population of multiple-planet systems
like these would be discovered by Kepler.

Kepler would go on to discover more than one hundred USP
planets and planet candidates, many in multiple planet systems
\citep[e.g.,][]{Batalha2011, Fressin2011, SanchisOjeda2013}.
A uniform analysis of the first 16 quarters of Kepler data by
\citet{SanchisOjeda2014} revealed that USP planet candidates have planet
radii $R_{\mathrm{p}} < 2~R_{\Earth}$ and occur around less than 1\%
of GK dwarfs.  The sub-day orbital periods of USP planets mean that they
should experience significant tidal interactions with their host stars.
As we will show in Section~\ref{section3}, if the efficiency of tidal
dissipation within the host stars of USP planets is the same as in hosts
of hot Jupiters, then USP planets should inspiral on a similar timescale.

There is reason to believe that the lower masses and shorter periods
of USP planets relative to hot Jupiters might affect the efficiency of
tidal dissipation within their host stars.  Detailed theoretical models of
nonlinear dissipation by internal gravity waves in stellar hosts indicate
that massive, short-period planets can trigger especially efficient
dissipation \citep[e.g.,][]{Barker2010, Essick2016}.  It may be that hot
Jupiters can trigger this mode of dissipation, while the lower-mass USP
planets cannot.  On the other hand, observations of hot Jupiter systems
have provided empirical evidence that shorter period systems experience
less efficient dissipation \citep{Penev2018}.  The extremely short
orbital periods of USP planets may therefore spare them from destruction.

The efficiency of tidal dissipation in USP planet host stars also plays
an important role in some models of their formation. \citet{Lee2017}
proposed a model that reproduced planet occurrence as a function of period
in which proto-USP planets form uniformly distributed in $\log_{10}(P)$.
USP planets are then brought to their observed locations by orbital decay
due to tidal dissipation within their host stars.  Other models do not
rely on tidal dissipation within host stars to explain the formation of
USP planets and instead appeal to tidal dissipation within the planet
\citep[e.g.,][]{Schlaufman2010, Petrovich2019}.

The efficiency of tidal dissipation inside USP planet host stars has
important consequences for both USP formation and evolution specifically,
as well as our understanding of tidal dissipation in general.  As shown in
\citet{Hamer2019}, systems hosting exoplanets destined to be destroyed by
tides will appear younger than a similar population of stars without such
planets.  In short, if USP planets are destroyed due to tides, then USP
planet host stars will be younger than similar stars without USP planets.

To evaluate their relative ages, in this paper we compare the Galactic
velocity dispersions of \citet{SanchisOjeda2014} USP planet candidate
host stars and stars without USP planets.  We show that these two
populations have indistinguishable kinematics and therefore similar ages.
This observation implies that USP planets do not experience tidal inspiral
during the main sequence lifetimes of their host stars.  The efficiency
of tidal dissipation inside a planet host star must therefore depend on
the amplitude and/or frequency of tidal forcing.  This paper is organized
as follows.  In Section 2, we describe our USP planet candidate host and
field star samples.  In Section 3, we outline our methods to make a robust
comparison between the Galactic velocity dispersions of the two samples.
In Section 4, we discuss the implications of our result for theories of
tidal dissipation and USP planet formation.  We conclude in Section 5.

\section{Data}\label{Section2}

We obtain our sample of USP planet candidate hosts from
\citet{SanchisOjeda2014}.  Those authors used Fourier-transformed Kepler
Q1--Q16 light curves to identify candidate transiting planets with $P<1$
day.  These systems were combined with additional candidate planets with
$P<1$ day from the KOI list as of January 2014 \citep{Kepler2014}, as
well as 28 candidates from other independent searches \citep{Ofir2013,
Huang2013, Jackson2013}.  These candidates were then vetted by a
homogeneous series of tests designed to identify false-positive signals.
This search resulted in a sample of 106 well-vetted USP planet candidates.

We obtain the Gaia Data Release 2 (DR2) designations of these USP
planet candidate host stars from SIMBAD and then query the Gaia Archive
to retrieve the astrometric and radial velocity data required to
calculate the kinematics of the sample.\footnote{For the details of
Gaia DR2 and its data processing, see \citet{GaiaMission,GaiaDR2},
\citet{GaiaCatalogValidation}, \citet{GaiaSpectrometer},
\citet{GaiaPhotValidation}, \citet{GaiaCalibration},
\citet{GaiaDR2RV}, \citet{Lindegren2018}, \citet{GaiaPhotProcessing},
\citet{GaiaSpectroProcessing}, and \citet{GaiaStandards}.} Most of the
stars have Gaia $G$-band magnitude $G\gtrsim16$, making them too faint to
have radial velocities available in Gaia DR2.  We obtain radial velocities
for these faint stars by supplementing our sample with radial velocities
(in order of priority) from the California-Kepler Survey \citep[CKS
-][]{CKS}, the Apache Point Observatory Galactic Evolution Experiment
\citep[APOGEE -][]{APOGEE} DR16, and the Large Sky Area Multi-Object Fiber
Spectroscopic Telescope (LAMOST) DR5 \citep{LAMOSTDR5}.  The majority of
the radial velocities for the USP planet candidate host stars come from
the CKS. We apply the data quality cuts described in \citet{Hamer2019}
and reproduced in the Appendix to ensure reliable kinematics.

We provide in Table~\ref{tab:tab1} the 68 USP planet candidate hosts
in our sample, their KIC identifiers, Gaia DR2 designations, radial
velocities, periods, radii, and masses.  Using the \citet{Brewer2018}
isochrone-derived values for host stellar radii $R_{\ast}$ and transit
depths from \citet{SanchisOjeda2014} we calculate $R_\mathrm{p}$.  We then
calculate planet mass $M_\mathrm{p}$ for the USP planet candidates in
our sample by fitting a spline to the Earth-like composition mass--radius
curve from \citet{Zeng2019}.

\begin{deluxetable*}{CCDCCC}
\tablecaption{Ultra-short-period Planet Hosts}
\tablenum{1}
\tablehead{\colhead{KIC ID} & \colhead{Gaia DR2 {\texttt source\_id}} & \twocolhead{Radial Velocity} &\colhead{Period} & \colhead{Planet Radius} & \colhead{Estimated Mass} \\
\colhead{} & \colhead{} & \twocolhead{(km s$^{-1}$)} & \colhead{(days)} & \colhead{($R_{\oplus}$)} & \colhead{($M_{\oplus}$)}}
    \decimals
    \startdata
6750902 &  2116704610985856512 & -17.00 &  0.469 &  2.579_{-0.105}^{+0.083} &  50.936_{-10.699}^{+10.927} \\
10186945 &  2119583510383666560 & -12.60 &  0.397 &  1.095_{-0.029}^{+0.017} &  1.394_{-0.133}^{+0.081} \\
10319385 &  2119593990103923840 & -38.40 &  0.689 &  1.611_{-0.031}^{+0.017} &  5.856_{-0.428}^{+0.244} \\
9873254 &  2119511080054847616 & -8.10 &  0.900 &  0.801_{-0.036}^{+0.035} &  0.444_{-0.068}^{+0.074} \\
6666233 &  2104748521545492864 & -51.43 &  0.512 & \cdots & \cdots \\
10647452 &  2107681262654003328 & -15.60 &  0.763 &  1.271_{-0.080}^{+0.056} &  2.409_{-0.514}^{+0.411} \\
5340878 &  2103579397088495616 & -10.90 &  0.540 & \cdots & \cdots \\
6755944 &  2104890633423618048 &  4.60 &  0.693 &  1.065_{-0.033}^{+0.029} &  1.256_{-0.136}^{+0.129} \\
5513012 &  2103628462794226304 & -11.60 &  0.679 &  1.523_{-0.037}^{+0.029} &  4.699_{-0.418}^{+0.355} \\
6265792 &  2103743018162573952 &  6.80 &  0.935 &  1.169_{-0.047}^{+0.043} &  1.776_{-0.251}^{+0.254} \\
    \enddata
    \tablecomments{Table 1 is ordered by right ascension and is published
    in its entirety in machine-readable format.  Planets without
    radius and mass estimates did not have their host stars' radii
    presented in \citet{Brewer2018}.}
    \label{tab:tab1}
\end{deluxetable*}

To evaluate the relative age of the USP planet candidate host population,
we need a sample of similar field stars with no detected USP planets.
As all of our USP planet candidate hosts lie in the Kepler field,
we select as our comparison sample all stars that were observed
for at least one quarter as part of Kepler's planet search program.
By selecting both samples from the Kepler field, we are ensuring that
the sample of stars not hosting planets has been thoroughly searched
for USP planets. \citet{SanchisOjeda2014} found that the occurrence
of USP planets is less than 1\% for GK dwarfs, so any contamination by
undetected USP planets should be minimal.  Additionally, as both samples
of stars are colocated in the Kepler field any kinematic differences
cannot be attributed to Galactic structure.  For these non-host stars,
we use SIMBAD to obtain their Gaia DR2 identifiers, and query the Gaia
Archive for their DR2 data.  As with the sample of USP planet candidate
host stars, many of the stars are too faint to have had their radial
velocities measured by Gaia.  We obtain radial velocities for these
stars with ground-based radial velocities from (in order of priority)
the CKS, APOGEE DR16, and LAMOST DR5.  Most of the radial velocities
for the field star sample come from LAMOST DR5.

\begin{figure}
    \centering
    \plotone{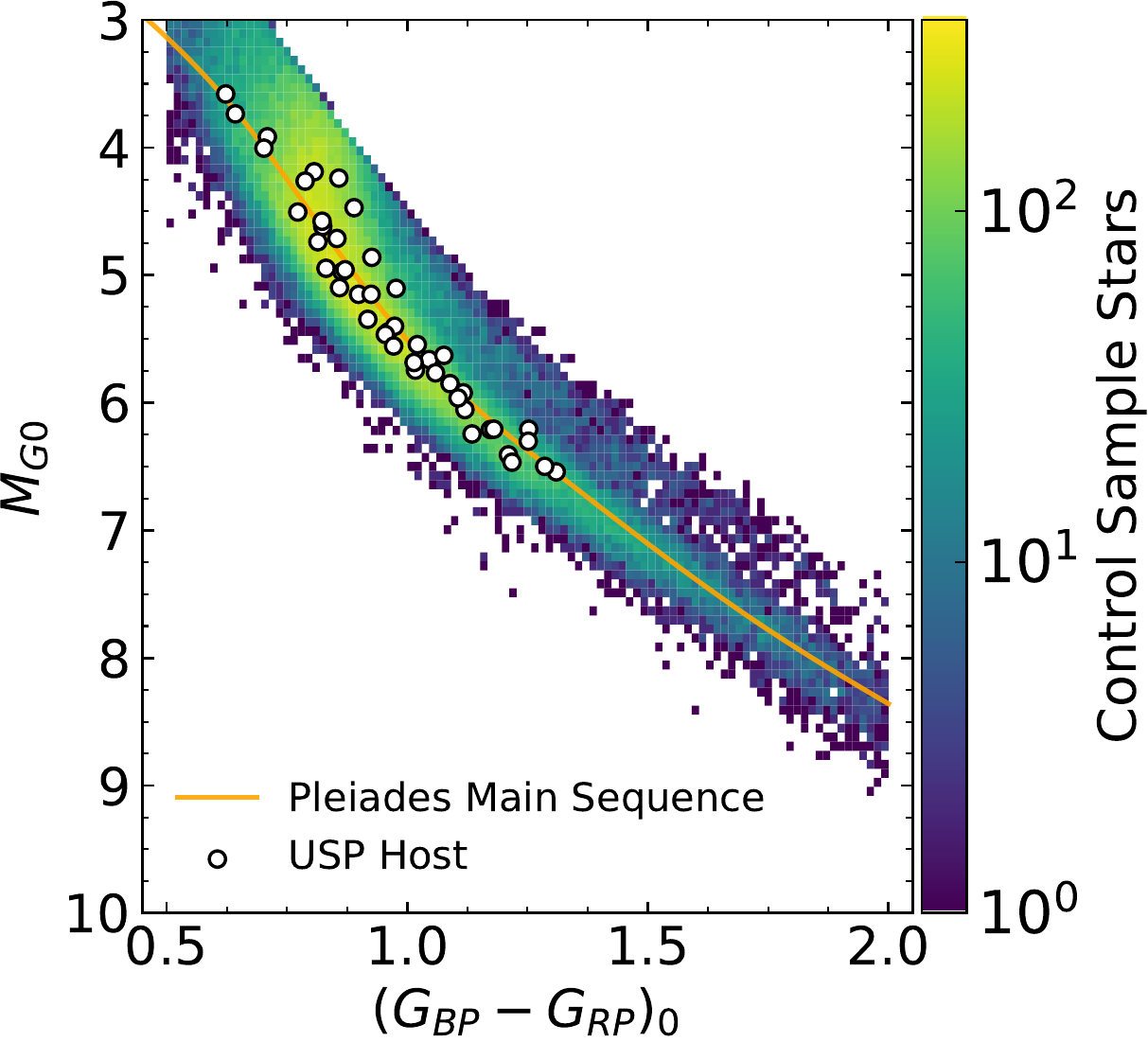}
    \caption{USP planet candidate host and field star samples.  We plot
    USP planet candidate hosts as white points with black outlines and
    the density of stars in the field star sample as the background
    color map.  We indicate the Pleiades main sequence spline fit from
    \citet{Hamer2019} used to remove evolved stars as the orange line.}
    \label{fig:Figure 1}
\end{figure}

To determine if USP planets tidally inspiral during the main sequence
lifetimes of their host stars, we must limit our sample of planet
candidate hosts and non-hosts to main sequence stars.  To do so,
we exclude all stars more than one magnitude above the Pleiades
solar metallicity zero-age main sequence from \citet{Hamer2019}.
Before applying this cut, we correct for extinction and reddening of
the stars in both samples using a three-dimensional extinction map
\citep{Capitanio2017}.  For a star in our sample, we interpolate the
grid of extinction values out to the star, and integrate along the
line-of-sight to calculate a total $E(B-V)$ reddening.  We convert
$E(B-V)$ to $E(G_\mathrm{BP}-G_\mathrm{RP})$ using the mean extinction
coefficients from \citet{Casagrande2018}.  We illustrate this calculation
in Figure~\ref{fig:Figure 1}.

\section{Analysis}\label{section3}

If one assumes that tidal dissipation within the host stars of USP planets
occurs with the same efficiency as in hot Jupiter hosts, then it can be
shown that USP planets should inspiral more quickly than hot Jupiters.
Assuming that all dissipation occurs in the host star---a safe assumption
for tidally locked planets---the inspiral time can \added{be }written
\begin{equation}\label{t_in}
    t_{\mathrm{in}} =  \frac{2}{13} \frac{2\Qs}{9}\frac{M_\ast}{M_{\mathrm{p}}}\left(\frac{a}{R_\ast}\right)^5\frac{P}{2\pi},
\end{equation}  
as in \citet{Hamer2019}.  Here $\Qs$ is the modified stellar tidal quality
factor, a parameter describing the efficiency of tidal dissipation,
$M_\ast$ is stellar mass, $a$ is the orbital semi-major axis, and $R_\ast$
is the stellar radius.  It follows that the ratio of the inspiral time
of a USP planet to that of a hot Jupiter around an identical star with
identical $\Qs$ is
\begin{equation}
    \frac{t_\mathrm{in, USP}}{t_\mathrm{in, HJ}}=\frac{M_\mathrm{HJ}}{M_\mathrm{USP}}\left(\frac{P_\mathrm{USP}}{P_\mathrm{HJ}}\right)^{13/3},
\end{equation}
by Kepler's third law.  The median period  of hot Jupiters in the
\citet{Hamer2019} sample was 3.4 days, whereas the median period of
USP planet candidates analyzed in this paper is 0.7 days.  Similarly,
the median mass of hot Jupiters in the \citet{Hamer2019} sample was
$290~M_{\oplus}$ while the median mass of USP planets analyzed in this
paper is $2.6~M_{\oplus}$.  As a result $t_\mathrm{in, USP}/t_\mathrm{in,
HJ}\approx0.10$.  Since \citet{Hamer2019} showed that hot Jupiters
inspiral during their host stars' main sequence lifetimes, if $\Qs$ is
the same for hot Jupiter and USP planet hosts then USP planets should
inspiral as well.

If this is so, then we should see a colder Galactic velocity dispersion
for USP planet host stars when compared to similar field stars.  To
calculate Galactic space velocities, we convert from the proper motions,
radial velocities, and parallaxes described in Section~\ref{Section2}
using \texttt{pyia} \citep{PriceWhelan2018}.  A requirement of this
approach is that the uncertainties on individual Galactic space velocities
are small relative to the velocity dispersion of the USP planet candidate
host star and field star samples.  We therefore estimate Galactic space
velocity uncertainties for each star using a Monte Carlo simulation.
We construct the astrometric covariance matrix and sample 100 realizations
from the astrometric uncertainty distributions for each star's position,
proper motions, parallax, and radial velocity using \texttt{pyia}.
The uncertainties on position, proper motion, and parallax all come from
Gaia DR2.  We source radial velocities and uncertainties from the CKS,
APOGEE DR16, Gaia DR2, and LAMOST DR5 in that order.  The typical radial
velocity precisions are 0.1, 0.5, 1.0, and 5.0 km s$^{-1}$ respectively.
We construct a point estimate for each star's velocity uncertainty by
taking the standard deviation of the 100 realizations of its Galactic
space velocity.

\begin{figure*}
    \centering
    \plotone{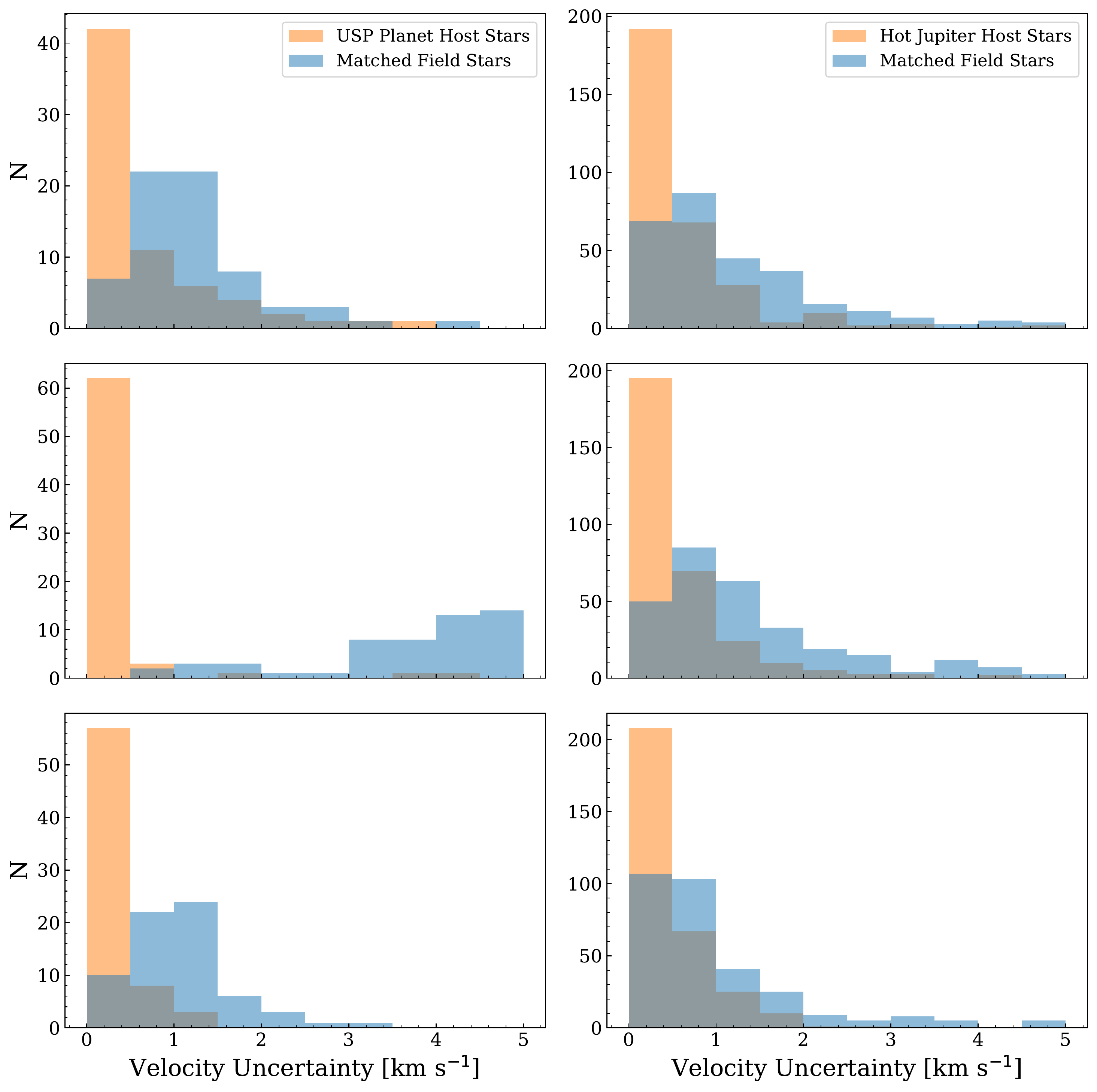}
    \caption{Distribution of $U$, $V$, and $W$ uncertainties in the top,
    middle, and bottom panels respectively.  Left: USP planet candidate
    host stars in orange and a matched control sample of field stars in
    blue.  Right: \citet{Hamer2019} hot Jupiter hosts in orange and a
    matched control sample of field stars in blue.  The typical velocity
    uncertainties are less than 1 km s$^{-1}$ for USP planet candidate
    host stars and less than 5 km s$^{-1}$ for the matched field star sample.
    Both are well below the population velocity dispersions.  We note
    that the typical velocity uncertainty in our sample of USP planet
    candidate host stars is smaller than the typical velocity uncertainty
    of the hot Jupiter host sample described in \citet{Hamer2019}.}
    \label{fig:Figure 2}
\end{figure*}

We plot in Figure~\ref{fig:Figure 2} the individual $UVW$ velocity
uncertainty distributions for both our USP planet candidate host star
sample and a matched control sample (the details of this matching is
described in the following paragraph).  Because our inference depends
on a comparison with the result from \citet{Hamer2019}, we execute
a similar calculation for the \citet{Hamer2019} sample.  We plot the
results of this calculation in Figure~\ref{fig:Figure 2}.  The typical
space velocities uncertainties are $\approx\!\!1$ km s$^{-1}$, much
smaller than the velocity dispersion of the stellar population (see
Figure~\ref{fig:Figure 3} below).  The uncertainty on $V$ for the field
star sample matched to the USP hosts is larger than the uncertainties
on $U$ and $W$.  The reason is that the Kepler field is aligned with
$V$ and the uncertainty is therefore dominated by the LAMOST radial
velocity uncertainties.  The typical velocity uncertainty for the USP
planet candidate host stars is smaller than that of the hot Jupiter host
star sample.

To perform a robust comparison of the kinematics of our USP planet
candidate host stars and field stars, we construct samples of field stars
matched to the USP planet candidate host stars on a one-to-one basis.
Since \citet{Winn2017} showed that USP planet candidate host stars
have a metallicity distribution indistinguishable from the field, we do
not attempt to match our samples on metallicity.  To mitigate possible
differences in stellar mass distributions, we assemble samples of field
stars matched to the sample of USP planet candidate hosts in color.
Specifically, we iteratively construct a color-matched control sample by
selecting 68 stars from the field star sample such that every USP planet
candidate host is mirrored by a star in the control sample within 0.025
mag in $(G_{BP}-G_{RP})_{0}$.  For each of these Monte Carlo iterations,
we calculate the mean $UVW$ velocity and then calculate the $UVW$
velocity dispersion
\begin{equation}
    \frac{1}{N}\sum \left[(U_i-\overline{U})^2+(V_i-\overline{V})^2+(W_i-\overline{W})^2\right]^{1/2}.
\end{equation}

We plot the result of this Monte Carlo simulation in Figure
\ref{fig:Figure 3}.  The USP planet candidate hosts have kinematics
indistinguishable from matched samples of non-host field stars.
As the Galactic velocity dispersion of a thin disk stellar population
is correlated with its average age \citep[e.g.][]{Binney2000}, the best
explanation for this observation is that USP planet candidate host stars
have ages consistent with the field.

\begin{figure*}
    \centering
    \plotone{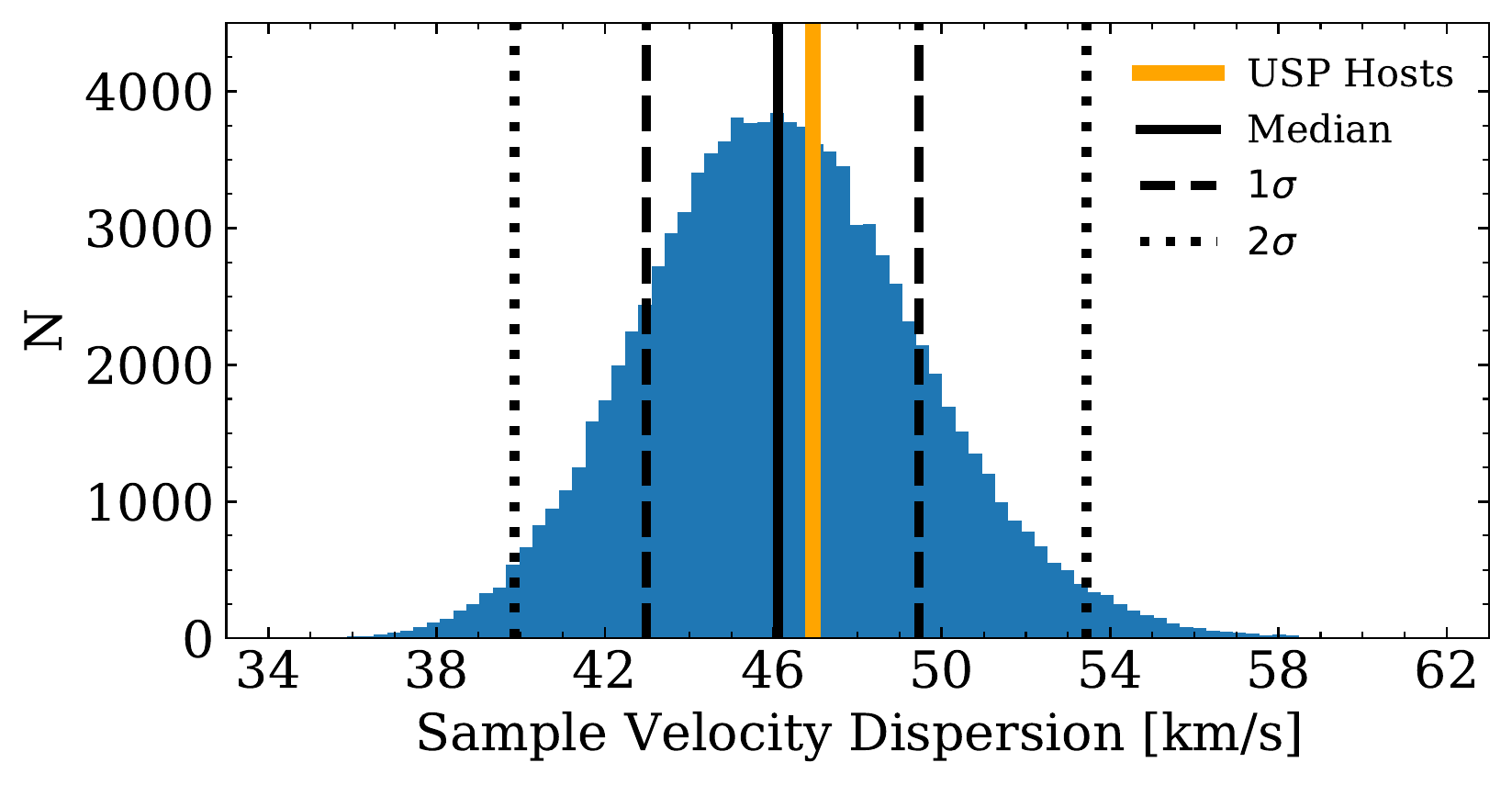}
    \caption{Velocity dispersion distribution of the matched control
    samples (blue histograms) compared to the velocity dispersion of
    our USP planet candidate host star sample (orange vertical line).
    The black vertical lines show the (2nd, 16th, 50th, 84th, and 98th)
    percentiles of the Monte Carlo samples.  The USP planet candidate
    hosts have kinematics indistinguishable from the matched Monte Carlo
    samples of field stars.  The only explanation for this observation is
    that USP planet candidate host stars are of similar age to similar
    non-host stars.  We argue that USP planets are stable against tidal
    inspiral during the main sequence lifetime of their host stars.}
    \label{fig:Figure 3}
\end{figure*}

\added{We have also ensured that our observation is not related to some
peculiarity of the Kepler field.  We first confirmed that the result
of \citet{Hamer2019} is robust when limiting the analysis to the Kepler
field. We compared the velocity dispersion of the 24 confirmed hot Jupiter
host stars in the Kepler field to similar stars without hot Jupiters using
the same algorithm including matching employed in \citet{Hamer2019}.
We find that confirmed hot Jupiter host stars in the Kepler field
have a colder velocity dispersion than matched samples of Kepler field
main sequence stars by about 2-$\sigma$.  The diminished significance
relative to \citet{Hamer2019} is a consequence of the much reduced hot
Jupiter host sample size in the Kepler field.  In addition, we checked
that the relatively warmer velocity dispersion of the USP planet hosts
in comparison to the all-sky hot Jupiter host sample is maintained in
the Kepler field. The 68 USP planet hosts have a velocity dispersion of
$46.95_{-0.14}^{+0.14}$ km s$^{-1}$, while the 24 hot Jupiter hosts in
the Kepler field have a velocity dispersion of $33.64_{-0.15}^{+0.15}$
km s$^{-1}$.  This velocity dispersion offset confirms our interpretation
that USP planets are robust to tidal evolution based on a comparison with
the larger all-sky hot Jupiter host sample analyzed in \citet{Hamer2019}.}

While we argue that our observation is evidence that USP planets do not
tidally inspiral, there are at least three other possible explanations
which must be ruled out.  Our observation could be attributed to a
large number of false positives in our USP planet candidate sample.
We believe this is unlikely.  \citet{SanchisOjeda2014} required each
transit be detected with SNR $>12$ and thoroughly vetted their USP
planet candidates with standard tests for false positives using Kepler
data \citep{Batalha2010}.  These included centroid shift checks to ensure
that the photocenter did not vary with the period of the candidate, which
would be indicative of a blended background eclipsing binary.  They also
searched for odd/even transit depth differences or phase-curve variations
indicative of eclipsing binaries.  As it has been shown that the false
positive rate in Kepler systems with multiple transiting planets is low
or even zero \citep[e.g.,][]{Lissauer2012, Lissauer2014}, the strongest
evidence that many of the USP planet candidates in our sample are real
is that 10 are found in multiple-planet systems.  Consequently, we argue
that it is highly unlikely that our sample has a high false positive rate.

Another possibility is that our sample size is too small to execute our
statistical comparison.  To verify that our sample size is sufficient,
we perform the following test.  As shown in Figure \ref{fig:Figure 3},
the Galactic velocity dispersion of the USP planet candidate host sample
is higher than 60\% of the matched Monte Carlo samples of field stars.
If the apparent velocity dispersion similarity is due to small number
statistics inflating the USP planet candidate host velocity dispersion,
then similarly-sized samples of hot Jupiter host stars would be affected
in the same way.  We select 1,000 random subsamples of 68 hot Jupiter
hosts from the \citet{Hamer2019} sample, construct 1,000 Monte Carlo
samples of field stars matched to each subsample of hot Jupiter hosts
as described in \citet{Hamer2019}, and determine how often the Galactic
velocity dispersion of the hot Jupiter host subsample is higher than
60\% of the matched Monte Carlo field star samples.  The result is
that identically zero of the 1,000 subsamples have a Galactic velocity
dispersion as relatively high as that of the USP planet candidate hosts.
As a result, we argue that there is less than a 1 in 1,000 chance that
the relatively small size of the USP planet candidate host star sample
affects our calculation.

It may also be that the difference in inspiral timescale between USP
planet candidate and hot Jupiter systems is due to differences in the
masses or radii of their host stars.  Using homogeneously derived stellar
parameters from \citet{Brewer2016} and \citet{Brewer2018}, we find that
the ratios of the median masses and radii of the USP planet candidate
and hot Jupiter host stars are 0.85 and 0.78.  Assuming similar host
star masses and radii implied that $t_\mathrm{in, USP}/t_\mathrm{in,
HJ}\approx0.1$.  After accounting for the difference in the median masses
and radii, $t_\mathrm{in, USP}/t_\mathrm{in, HJ}$ increases to 0.29.
While the median USP planet candidate and hot Jupiter host stars differ,
USP planets should still inspiral on a shorter timescale than hot Jupiters
if $\Qs$ is independent of forcing frequency and/or amplitude.

\section{Discussion}\label{section4}

We have shown that main sequence stars hosting USP planet candidates have
a Galactic velocity dispersion indistinguishable from that of matched
samples of stars which do not host observed USP planets.  This implies
that the populations have similar ages and that USP planets do not
tidally inspiral during the main sequence lifetimes of their host stars.
This is in sharp contrast to hot Jupiters, which have been shown to
tidally inspiral on this timescale \citep{Hamer2019, Yee2020}.  As we
argued above, there are no other plausible explanations for the similar
kinematics of USP planet candidate hosts and non-hosts other than the
robustness of USP planets to tidal inspiral.  This requires that USP
planets trigger less efficient dissipation within their host stars than
hot Jupiters.

One possible explanation for this change in efficiency may be that
$\Qs$ is a function of tidal forcing frequency.  In this case, the
shorter orbital periods of USP planets in comparison to hot Jupiters
could be the key to their survival.  There are both theoretical reasons
\citep[e.g.][]{Ogilvie2012, Duguid2020} to believe that this might be
so and some observational evidence that $\Qs$ increases as orbital
period decreases.  \citet{Penev2018} compared the rotation rates of
stars with $T_\mathrm{eff}<6100$ K hosting hot Jupiters with $P<3.5$ days
to the expected rotation rates for similar stars without hot Jupiters.
They then determined the efficiency of tidal dissipation within the host
stars necessary to explain the observed rotational enhancements over
the systems' lifetimes.  They found that $\Qs$ increases from $10^5$
to $10^7$ as the tidal period
\begin{equation}
    P_\mathrm{tide}=\frac{1}{2(P_\mathrm{orb}^{-1}-P_\mathrm{spin}^{-1})},
\end{equation} decreases from 2 days to 0.5 days.  This result is
consistent with our inference that the tidal dissipation triggered by hot
Jupiters within their host stars is more efficient than that triggered
by the shorter-period USP planets.

USP planets are also two orders of magnitude less massive that hot
Jupiters.  Theoretical work on non-linear internal gravity waves has
shown that the efficiency of tidal dissipation within stars may depend
on the amplitude of tidal forcing \citep{Barker2010, Essick2016}.
According to \citet{Barker2010}, the non-linear wave breaking criterion is
\begin{equation}
\label{nonlinearity}
    \left(\frac{M_\mathrm{p}}{M_\mathrm{Jup}}\right)\left(\frac{P}{1\mathrm{\ day}}\right)^{1/6}\gtrsim3.3.
\end{equation}

As described above, the median orbital period of hot Jupiters in the
\citet{Hamer2019} sample was 3.4 days, whereas the median orbital period
of USP planet candidates analyzed in this paper is 0.7 days.  Because
Equation~\eqref{nonlinearity} depends on orbital period, non-linear wave
breaking might be important for planets with $M_{\mathrm{p}} \gtrsim
2.7~M_{\mathrm{Jup}}$ and $M_{\mathrm{p}} \gtrsim 3.5~M_{\mathrm{Jup}}$
at $P=3.4$ days and $P=0.7$ days.  Only 44 out of 313 hot Jupiters
and zero USP planet candidates satisfy Equation~\eqref{nonlinearity}.
Therefore, the \citet{Barker2010} model cannot explain the apparent
difference in the efficiency of tidal dissipation we infer between hot
Jupiter and USP planet systems.

\citet{Essick2016} proposed that weakly non-linear gravity waves could
result in amplitude-dependent $\Qs$ values.  Those authors presented a
numerical fit to their predicted tidal inspiral times which is valid
for $0.5~M_\mathrm{Jup} \leq M_\mathrm{p} \leq 3.0~M_\mathrm{Jup}$
and $P<4$ days.  None of the USP planet candidates in our sample are
massive enough to trigger this mode of dissipation.  Of the 50 hot
Jupiters in the \citet{Hamer2019} sample for which the numerical fit
is valid and for which we have homogeneously derived stellar parameters
from \citet{Brewer2016} and \citet{Brewer2018}, 16 have tidal inspiral
times shorter than the main sequence lifetimes of their host stars.
While weakly non-linear internal gravity waves may be capable of
explaining the inspiral of a minority of hot Jupiter systems, they cannot
explain the observation that most hot Jupiters do not survive their host
stars' main sequence lifetimes.  The net result is that we can rule out
weakly non-linear internal gravity waves as a likely explanation for
the difference we infer in tidal dissipation efficiency between the hot
Jupiter and USP planet regimes.

As in \citet{Hamer2019}, we can derive a limit on the stellar tidal
quality factor $\Qs$ based on our observation that USP planets do not
tidally inspiral during the main sequence lifetimes of their host stars.
Using homogeneously-derived stellar parameters from \citet{Brewer2018},
we calculate the main sequence lifetime of each USP planet candidate
host star according to the scaling relation
\begin{equation}\label{ms_scale}
    \frac{t_{\mathrm{MS,}\ast}}{t_{\mathrm{MS,}\odot}}=\left(\frac{M_{\ast}}{M_{\odot}}\right)^{-2.5}.
\end{equation}

\noindent
Finally, we solve Equation~\eqref{t_in} for $\Qs$, assuming
$t_{\mathrm{in}} > t_{\mathrm{MS}}$ to obtain a lower limit on $\Qs$
for each system
\begin{equation}\label{Q_ul}
    \Qs > t_{\mathrm{MS}}\frac{117}{4}\frac{M_{\mathrm{p}}}{M_{\ast}}\left(\frac{R_{\ast}}{a}\right)^5\frac{2\pi}{P}.
\end{equation}

\begin{figure}
    \centering
    \plotone{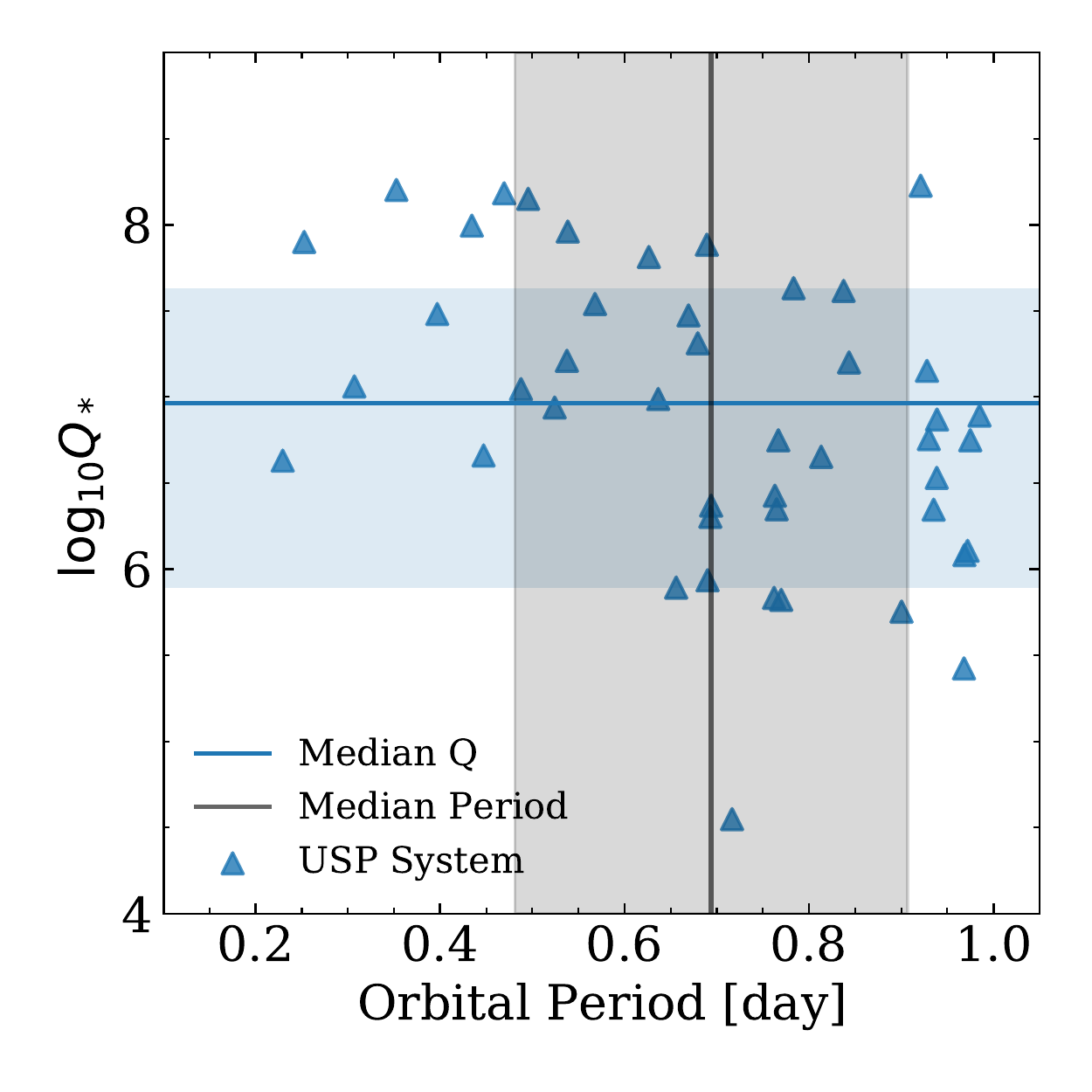}
    \caption{Minimum $\Qs$ required for survival of USP planets during
    the main sequence.  We use the formalism presented in \citet{Lai2012}
    and homogeneously derived spectroscopic stellar parameters from
    \citet{Brewer2018} to derive the limit on $\Qs$.  The vertical line
    shows the median period of the USP planet candidate sample, whereas
    the gray rectangle spans the 16th to 84th percentiles of the sample
    period distribution (approximately 0.39 days $\lesssim P \lesssim$
    0.92 days).  We calculate the minimum $\Qs$ required for each system,
    and within this period range where our velocity dispersion analysis
    applies, we calculate the (16,50,84) percentiles.  For those systems
    in the gray rectangle, the horizontal line is the median $\Qs$
    and the blue rectangle spans the 16th to 84th percentiles of the
    inferred $\Qs$ distribution.  We find $\logQs> 6.96^{+0.67}_{-1.07}$}
    \label{fig:Q_LL}
\end{figure}

We plot the results of this calculation in Figure~\ref{fig:Q_LL}.
Because we use a population-level approach, we can only provide
constraints based on the ``typical'' system within our sample.
We estimate $\Qs$ in the typical USP planet system by calculating the
median $\Qs$ among the systems with periods that fall within the 16th
and 84th period percentiles (instead considering the typical USP planet
in terms of mass rather than period makes a negligible difference).
We find that the survival of USP planets beyond the end of their hosts'
main sequence lifetimes requires $\logQs > 6.96^{+0.67}_{-1.07}$.

Our observation that USP planets are robust to tidal dissipation inside
their host stars also informs theories of USP planet formation.  In the
USP planet formation scenarios put forward by \citet{Schlaufman2010}
and \citet{Petrovich2019}, non-zero eccentricities of proto-USP planets
are maintained due to secular interactions with more distant planets in
multiple-planet systems.  The \citet{Schlaufman2010} scenario suggests
maximum eccentricities of proto-USP planets $e \lesssim 0.1$ while
the \citet{Petrovich2019} scenario proposes maximum eccentricities
$e \gtrsim 0.1$.  In both cases, USP planets arrive at their current
orbits as their orbits circularize due to tidal dissipation in the USP
planets themselves.  In contrast, the model favored by \citet{Lee2017}
assumes that planets with $M_{\mathrm{p}} \approx 5~M_{\oplus}$ formed
with a uniform distribution in $\log_{10}(P)$ from the inner edge of the
protoplanetary disk thought to be corotating with the star at $P\approx1$
day to $P=400$ days.  After the era of planet formation, tidal dissipation
within the host stars removes orbital energy and angular momentum from
the proto-USP planets and brings them to their observed orbital periods.
Those authors found that $\Qs\sim10^7$ acting over 5 Gyr best reproduced
the occurrence of USP planets as a function of period.

To determine if the \citet{Lee2017} scenario is consistent with
our observation and the detailed USP planet host star data from
\citet{Brewer2018}, we integrate the orbits of the USP planet candidates
in our sample backward in time over 5 Gyr with $\Qs=10^7$ according to
Equation (5) of \citet{Lee2017}.  We find that none of the USP planet
candidates in our sample could have migrated from an initial $P>1$ day
due to tidal dissipation inside their host stars.  Therefore we argue
that it is unlikely that tidal dissipation within host stars plays an
important role in the formation of USP planets.  Even the possible period
dependence of $\Qs$ proposed in \citet{Penev2018} would not allow for
significantly greater migration of proto-USP planets.   In that scenario,
$\Qs$ decreases as tidal period increases from $P_\mathrm{tide}\approx0.5$
day to $P_\mathrm{tide}\approx2$ days.  As main sequence FGK dwarfs
typically have rotational periods of at least 5--10 days by 1 Gyr
\citep[e.g.][]{Rebull2018, Rebull2020},  $P_\mathrm{tide}=0.5$ days
corresponds to an orbital period of 0.83--0.91 days.  Only 30\% of the
USP planet candidates in our sample could have migrated from beyond 0.83
days, so the majority could not have migrated from the range of tidal
period where $\Qs$ begins to decrease.

\added{If USP planets take a few Gyr to arrive at their present locations
as in the \citet{Lee2017} scenario, then our observation that USP planet
hosts have an age consistent with the field may not say anything about
tidal evolution.  Alternatively, the USP formation models proposed by
\citet{Schlaufman2010} and \citet{Petrovich2019} imply the early arrival
of USP planets at their observed locations via eccentricity excitation
and circularization. In these models, the overall timescale for this
eccentricity driven migration is the sum of multiple cycles of secular
eccentricity excitation and tidal damping.  In the \citet{Schlaufman2010}
scenario, many cycles involving only a small eccentricity excitation
are required.  On the other hand, in the \citet{Petrovich2019} scenario
only a few cycles with larger eccentricity excitation are necessary.
Consequently, the migration time in the \citet{Schlaufman2010} USP
planet formation model will be longer than in the \citet{Petrovich2019}
model.}

\added{To estimate the overall eccentricity driven migration time in
the \citet{Schlaufman2010} scenario, we imagine a proto-USP planet with
$M_{\mathrm{p}} = 2.6~M_{\oplus}$ and $R_{\mathrm{p}} = 1.3~R_{\oplus}$
initially orbiting at $P = 2 \mathrm{~days} \Leftrightarrow a = 0.03$ AU
a star with $M_{\ast} = 0.87~M_{\odot}$.\footnote{The planet mass, planet
radius, and stellar mass are median values for the systems in our sample.}
We simulate cycles of eccentricity excitation followed by subsequent tidal
circularization for this proto-USP planet.  We assume an eccentricity
excitation of 0.001 on each cycle on a timescale corresponding to $10^4$
orbits of an external planet with $P = 10$ days.  We then estimate the
circularization time according to Equation~2 of \citet{Mardling2007}
assuming the proto-USP planet has tidal parameters similar to those
for the Earth given in \citet{Goldreich1966}.  The elapsed time in
a single cycle is therefore the sum of the eccentricity excitation
and circularization timescales.  We update the orbit according to the
relation $a_{i+1} = a_{i} (1-e^2)$ and count the number of iterations
and total elapsed time required to migrate the proto-USP planet from
0.03 AU to 0.017 AU (the median semimajor axis of the USP planets in
our sample).  We find that nearly $6 \times 10^{5}$ cycles over 160 Myr
are sufficient to move the proto-USP planet to the median location of our
sample of USP planets.  The eccentricity driven migration timescale in
the \citet{Petrovich2019} scenario will be even shorter.  This timescale
of 160 Myr is much smaller than the main sequence lifetimes of the
stars searched for USP planets by Kepler, so the observation that USP
planet hosts have a population age consistent with similar field stars
implies that USP planets are stable against tidal decay.} 
We conclude
that models of USP planet formation that invoke tidal dissipation within
the USP planet itself as it circularizes are the most likely explanation
for the origin of USP planets as a class \citep[e.g.,][]{Schlaufman2010,
Petrovich2019}.

\section{Conclusion}

We compare the kinematics of USP planet candidate hosts to matched
samples of stars without observed USP planets using data from Gaia
DR2 supplemented by ground-based radial velocities from a variety
of sources.  If tidal dissipation inside USP planet host stars is
similarly efficient to that in hot Jupiter host stars, then USP planets
should tidally inspiral during their host stars' main sequence lifetimes
like hot Jupiters.  If this is so, then stars that are observed to host
USP planets should be systematically younger than similar field stars.
On the other hand, the observation that USP planet hosts have similar
ages to field stars would imply the robustness of USP planets to tidal
dissipation and would support theoretical models of tidal dissipation
inside of exoplanet host stars that suggest tidal dissipation depends on
forcing frequency and/or amplitude.  We find that USP planet candidate
host stars have a similar Galactic velocity dispersion and therefore a
population age consistent with matched samples of field stars without
observed USP planets.  The implication is that unlike hot Jupiters,
USP planets do not inspiral during the main sequence lifetimes of their
host stars.  We find that $\Qs\gtrsim10^7$ at $P \approx 0.7$ day and
$M_{\mathrm{p}} \approx 2.6~M_{\oplus}$.  We argue that this observation
supports models of tidal dissipation in which the efficiency of tidal
dissipation in the host star depends on the amplitude and/or frequency
of tidal forcing in the range $0.5 \mathrm{~days} \lesssim P \lesssim 5$
days and $1~M_{\oplus} \lesssim M_{\mathrm{p}} \lesssim 1000~M_{\oplus}$.
We propose that the observed USP planet population is best explained by
scenarios of USP planet formation which rely on tidal dissipation within
the USP planet itself due to eccentricity excitation and subsequent
circularization.

\acknowledgements
We thank Adrian Barker, Kaloyan Penev, Cristobal Petrovich, and Nevin Weinberg
for helpful comments on this paper.  This paper includes data
collected by the Kepler mission.  Funding for the Kepler mission
is provided by the NASA Science Mission directorate.  This work
has made use of data from the European Space Agency (ESA) mission
{\it Gaia} (\url{https://www.cosmos.esa.int/gaia}), processed
by the {\it Gaia} Data Processing and Analysis Consortium (DPAC,
\url{https://www.cosmos.esa.int/web/gaia/dpac/consortium}).  Funding for
the DPAC has been provided by national institutions, in particular the
institutions participating in the {\it Gaia} Multilateral Agreement.
This research has made use of the NASA Exoplanet Archive, which is
operated by the California Institute of Technology under contract with
NASA under the Exoplanet Exploration Program.  This research has made use
of the SIMBAD database, operated at Centre de Donn\'ees astronomiques
de Strasbourg (CDS), Strasbourg, France \citep{SIMBAD}.  This research
has made use of the VizieR catalog access tool, CDS, Strasbourg, France
(DOI: \href{http://doi.org/10.26093/cds/vizier}{10.26093/cds/vizier}).
The original description of the VizieR service was published in
A\&AS 143, 23 \citep{VizieR}.  This research made use of Astropy
(\url{http://www.astropy.org}) a community-developed core Python package
for Astronomy \citep{astropy:2013, astropy:2018}.

\vspace{5mm}
\facilities{Exoplanet Archive, Gaia, Kepler}

\software{\texttt{Astropy} \citep{astropy:2013, astropy:2018},
          \texttt{pyia} \citep{PriceWhelan2018},
          \texttt{pandas} \citep{pandas}
          }

\clearpage
\bibliography{mybib_v2}

\appendix\label{app1}

\noindent
\citet{Lindegren2018} and \cite{Marchetti2019} suggest the
following quality cuts to ensure reliable astrometry.  We apply
them to our field star sample.  Cuts 4 and 8 are cuts C.1 and C.2 of
\citet{Lindegren2018}, where $u$ is the unit weight error and $E$ is
the $\texttt{phot\_bp\_rp\_excess\_factor}$.  Both cuts are related
to problems that arise due to crowding.  Cut C.1 removes sources for
which the single-star parallax model does not fit well, as two nearby
objects are instead mistaken for one object with a large parallax.
Cut C.2 removes faint objects in crowded regions, for which there are
significant photometric errors in the $G_{BP}$ and $G_{RP}$ magnitudes.
We also impose astrometric quality cuts 1-4 to the USP planet candidate
host sample.  We do not apply cut 6 to the USP planet candidate host
sample because it is known that the reflex motion of planet host stars
can result in excess noise in the astrometric fitting \citep{Evans2018}.
We do not apply cut 7 to the USP planet candidate host sample because
the many of the USP planet candidate host radial velocities come from
ground-based radial velocities.  Overall, these cuts are designed to
produce a sample with high-quality astrometry.

\begin{enumerate}
    \item $\texttt{parallax\_over\_error} > 10$
    \item $-0.23 < \texttt{mean\_varpi\_factor} < 0.36$
    \item $\texttt{visibility\_periods\_used} > 8$
    \item $u < 1.2*\texttt{MAX}(1,\exp{(-0.2*\texttt{phot\_g\_mean\_mag}-19.5}))$
    \item $\texttt{astrometric\_gof\_al} <3$
    \item $\texttt{astrometric\_excess\_noise\_sig}<2$
    \item $\texttt{rv\_nb\_transits}>5$
    \item $1.0+0.0015*\texttt{bp\_rp}^2 < E < 1.3+0.06*\texttt{bp\_rp}^2$
    \item $\texttt{phot\_bp\_mean\_flux\_over\_error} >10$
    \item $\texttt{phot\_rp\_mean\_flux\_over\_error} >10$
\end{enumerate}

\listofchanges 

\end{document}